\documentclass{elsart}%
\usepackage{amsfonts}
\usepackage{amsmath}
\usepackage{amssymb}
\usepackage{graphicx}%
\providecommand{\U}[1]{\protect\rule{.1in}{.1in}}
\begin{document}
%
\begin{frontmatter}%
%

\title
{Structural dynamics in thermal-treatment of amorphous indium-oxide films}%
%

\author{Z. Ovadyahu}%
%

\address
{Racah Institute of Physics, The Hebrew University, Jerusalem 91904, Israel}%
%

\begin{abstract}
Thermally-treating amorphous indium-oxide films is used in various basic
studies as a means of tuning the system disorder. In this process the
resistance of a given sample decreases while its amorphous structure and
chemical composition is preserved. The main effect of the process is an
increase in the system density which in turn leads to improved interatomic
overlap which is easily detected as improved conductivity. A similar effect
has been observed in studies of other amorphous systems that were subjected to
pressure. In the current work we show that the Raman spectra of amorphous
indium-oxide change in response to thermal-treatment in a similar way as in
pressure experiments performed on other disordered and amorphous systems. We
present a study of how thermal-treatment changes the system dynamics by
monitoring the resistance versus time of indium-oxide films following various
stages of thermal-treatment. The time dependence of the sample resistance fits
the stretched exponential law with parameters that change systematically with
further annealing. Implication of these results to slow dynamics phenomena
that are governed by the Kohlrausch's law are discussed.
\end{abstract}%
%

%

\end{frontmatter}%

\subsection{Introduction}

Understanding the properties of amorphous solids has been a great intellectual
endeavor. The lack of long-range periodicity presents a challenge to theory,
and the non-equilibrium nature of the systems adds another layer of
complexity. In addition to lack of long-range order, most amorphous solids
exhibit mass density that is lower than their crystalline counterpart. This is
presumably due to their being formed by a fast cooling from the liquid or
gaseous phase. The latter preparation method, quench-cooling the material from
the vapor phase onto a cold substrate, usually results in a spongy structure
that has many micro-voids reducing the material bulk specific gravity.
Consequently, an appreciable volume change may be affected in these structures
upon application of pressure. Indeed, amorphous systems often show
considerable volume-change $\Delta$V under pressure; relative volume shrinkage
$\Delta$V/V exceeding 15\% was observed in a number of studies
\cite{1,2,3,4,5,6,7,8,9,10,11,12,13}.

It was recently shown \cite{14} that similar densification may be affected by
thermally-treating amorphous indium-oxide In$_{\text{x}}$O films. The protocol
used a cycle of heating-cooling on vapor-deposited films and demonstrated a
volume change of up to $\simeq$20\%. At the same time, a change of
several\textit{ }orders of magnitude of the film resistance at
room-temperature was observed in this low carrier-concentration version of
In$_{\text{x}}$O \cite{15}.

In addition to the similarity between the effects produced by pressure and
thermal-treatment, in terms of densification, both protocols seem to share a
peculiar after-effect; upon relief of pressure, and after cooling-back to
room-temperature in the thermal-treatment protocol, the system volume slowly
crept back up towards its initial value. This "swelling" effect was observed
in the time dependence of the optical-gap of glasses following pressure relief
\cite{7,14}. In our previous study of In$_{\text{x}}$O films the effect was
observed in the time-dependence of the optical-gap and the system resistance
studied \textit{in-situ} after the thermal-treatment is terminated \cite{14}.

Previous thermal-treatment experiments \cite{14} were performed using versions
of In$_{\text{x}}$O with relatively small deviations from the In$_{\text{2}}%
$O$_{\text{3}}$ stoichiometry of the ionic compound. In this work we use
indium-richer version of In$_{\text{x}}$O with larger carrier-concentration
\textit{N.} This version is commonly used in superconductor-insulator
transitions studies \cite{16}. It exhibits the main features of all
In$_{\text{x}}$O versions in having the flexibility of being able to modify
its volume by heat-treatment and thus fine-tune its resistance while the
relatively high carrier-concentration (\textit{N} $\gtrsim$5x10$^{\text{20}}%
$cm$^{\text{-3}}$). This makes it possible to observe superconductivity at
liquid-helium temperatures once the system disorder is low enough. The
In$_{\text{x}}$O system affords a wide range of carrier-concentration (while
keeping the Fermi energy in the conduction-band) with \textit{N} as low as
8x10$^{\text{18}}$cm$^{\text{-3}}$ to \textit{N}$\approx$10$^{\text{22}}%
$cm$^{\text{-3}}$. The carrier-concentration of In$_{\text{x}}$O is
essentially determined by the In/O ratio during material deposition \cite{17}.
Due to its high carrier-concentration, the as-prepared film of the material
naturally has much lower resistivity than In$_{\text{x}}$O films with
\textit{N}$<$10$^{\text{20}}$cm$^{\text{-3}}$. The amount by which the
room-temperature resistivity of the sample could be reduced by thermal
treatment is accordingly more modest; it is rarely more than a factor of
$\approx$4. On the other hand the accompanying relative change in other
physical properties, per change in resistivity, is actually larger \cite{15}.

Taking advantage of this sensitivity and the large response this material
exhibits in Raman scattering, we study how the boson-peak changes with
response to the thermal-treatment procedure used in taming the system
disorder. The boson-peak is a feature that appears in Raman spectra of
disordered systems as a peculiar broad peak at 10-100 cm$^{\text{-1}}%
$\cite{18,19,20,21,22}. The magnitude of the boson-peak, characteristically
observed in glasses, decreases with pressure \cite{22}. As it will be shown
below, the same occurs upon thermally treating In$_{\text{x}}$O samples
further supporting the similarity between applying pressure and thermal
treatment of glasses \cite{14}. We note however that the time dependence of
the Raman spectra while pressure is applied or after its release was not yet
studied. To our knowledge this aspect has not been studied in any glassy
system probably due to technical difficulty. By contrast, monitoring the
resistance is an extremely sensitive measure of the structural change
associated with densification of In$_{\text{x}}$O as demonstrated in
\cite{14,15}. Resistance measurement is also a fast process which allows
tracking these changes in real time.

A special emphasis is given in this work to the `after-effect' mentioned above
where the resistance of the samples slowly increases after the heat-treatment
has ended and the sample temperature returned to its pre-treatment value. The
behavior of the resistance during this period has a mechanical analogue; The
volume of a `memory-foam' that has endured the squashing effect of a heavy
object will show qualitatively similar time-dependence as R(t) after the heat
is turned off. In particular, it will swell back towards its original volume,
partly as a rather fast change when the weight is lifted followed by a much
slower process.

The time dependence of the swelling phenomenon, as reflected in the sample
resistance, fits a stretched-exponential law. A systematic study of how this
law changes with consecutive annealing-cycles reveals that the in addition to
densification, the rate distribution that determines the slow structural
dynamics becomes narrower. The implication of these results for data analysis
of glass dynamics and transport studies are discussed.

\subsection{Samples preparation and characterization}

The In$_{\text{x}}$O films used here were e-gun evaporated onto
room-temperature substrates using 99.999\% pure In$_{\text{2}}$O$_{\text{3}}$
sputtering-target. Undoped Silicon wafers were used as substrates for both,
electrical measurements and Raman spectroscopy measurements. Deposition was
carried out at the ambience of 3$\pm$0.5x10$^{\text{-5}}$ Torr oxygen pressure
maintained by leaking 99.9\% pure O$_{\text{2}}$ through a needle valve into
the vacuum chamber (base pressure $\simeq$10$^{\text{-6}}$ Torr). Rates of
deposition were 1.4-2.5~\AA /s. With this range of rate-to-oxygen-pressure,
the In$_{\text{x}}$O samples had carrier-concentration \textit{N} in the range
(6-25)x10$^{\text{21}}$cm$^{\text{-3}}$ measured by Hall-Effect at
room-temperature. Note that the carrier-concentration of the samples used in
the previous study was considerably lower \cite{14}.The evaporation source to
substrate distance in the deposition chamber was 45cm. This yielded films with
thickness uniformity of $\pm$2\% across a 2x2cm$^{\text{2}}$ area. Lateral
sizes of samples used for transport measurements was typically
1x2mm$^{\text{2}}$ (width x length respectively), and 1x1cm$^{\text{2}}$ for
the Raman spectroscopy. The films thickness of the samples used for transport
measurements was 510$\pm$10 \AA .

The as-deposited samples typically had sheet-resistance R$_{\square}$ of the
order of $\approx$(5-10)x10$^{\text{3}}\Omega$ at room-temperatures, much
smaller than the films with the lower carrier-concentrations used before. This
was usually the starting stage for the thermal-treatment cycles performed on
each preparation batch (3 different batches were used in the study). A
description of the annealing process will be described in the next section.
Details of the changes in the material microstructure in the process of
thermal-treatment are described elsewhere \cite{14,15}.

\subsection{Measurements methods}

After removal from the deposition chamber, the sample was mounted onto a
heat-stage in a small vacuum cell wired to make contacts with the sample for
electrical measurements and a thermocouple thermometer attached to the
sample-stage. The cell used for monitoring the sample resistance and
temperature had a light-weight made of 0.2mm copper-sheet equipped with a
thermofoil\texttrademark\ heating-strip on its back side. The characteristic
time to reach 90\% of the asymptotic temperature after applying power to the
heating-element was typically $\approx$300s for the range of temperatures used
in this work.

Copper wires were soldered to indium contacts pressed into the sample strip to
facilitate resistance measurements. These were performed by a two-terminal
technique using either the computer-controlled HP34410A multimeter or the
Keithley K617.

Raman spectra were taken with a Renishaw inVia Reflex Spectrometer using a
laser beam with either 514 nm or 785 nm wavelength and edge-filter at
$\approx$70cm$^{\text{-1}}$.

\section{Results and discussion}

\subsection{The thermal treatment protocol}

The protocol for thermally-treating In$_{\text{x}}$O samples is composed of
the following steps: The sample, deposited at T$_{\text{P}}$ (typically$\simeq
$298$\pm$2K) is anchored to a heat-stage within the measuring cell and
electrical contacts are made to allow its resistance to be monitored. The cell
is then evacuated by a rotary-pump to a pressure of $\lesssim$0.03mbar. Next,
the heating-stage is energized, and within a time interval $\delta$t (of the
order of $\approx$300s) the sample-stage reaches an annealing temperature
T$_{\text{A}}$. The system is then kept at this temperature for a dwell-time
t$_{\text{d}}$, typically much longer than $\delta$t. Finally, the heat supply
is turned off and the sample is cooled back to ambient temperature within
essentially the same $\delta$t as in the heat-up stage. The sample resistance
and the stage temperature are continuously measured throughout the
annealing-cycle. A typical set of results for an annealing cycle is
illustrated in Fig.1a including both R(T) and $\Delta$T(t)$\equiv$%
T$_{\text{A}}$(t)-T$_{\text{RT}}$(t) where~T$_{\text{RT}}$ is the ambient
temperature. The sharp response of the resistance while the temperature
increases upon turning on the heating and during the cool-back to
T$_{\text{RT}}$ is due to the temperature dependence of the sample resistivity
(the temperature coefficient of the resistance is negative for all the samples
studied here). Note however that the resistance changes even during the parts
of the cycle where the temperature is essentially constant (Fig.1b and
Fig.1c). In particular, the resistance keeps going up long after the stage
temperature stabilizes (Fig.1b). The time of R is associated with the increase
in the sample volume. This "swelling" effect was also observed in
pressure-densification studies after the applied pressure has been relieved
\cite{6,7}.%
\begin{figure}[ptb]%
\centering
\includegraphics[
height=3.9055in,
width=5.5486in
]%
{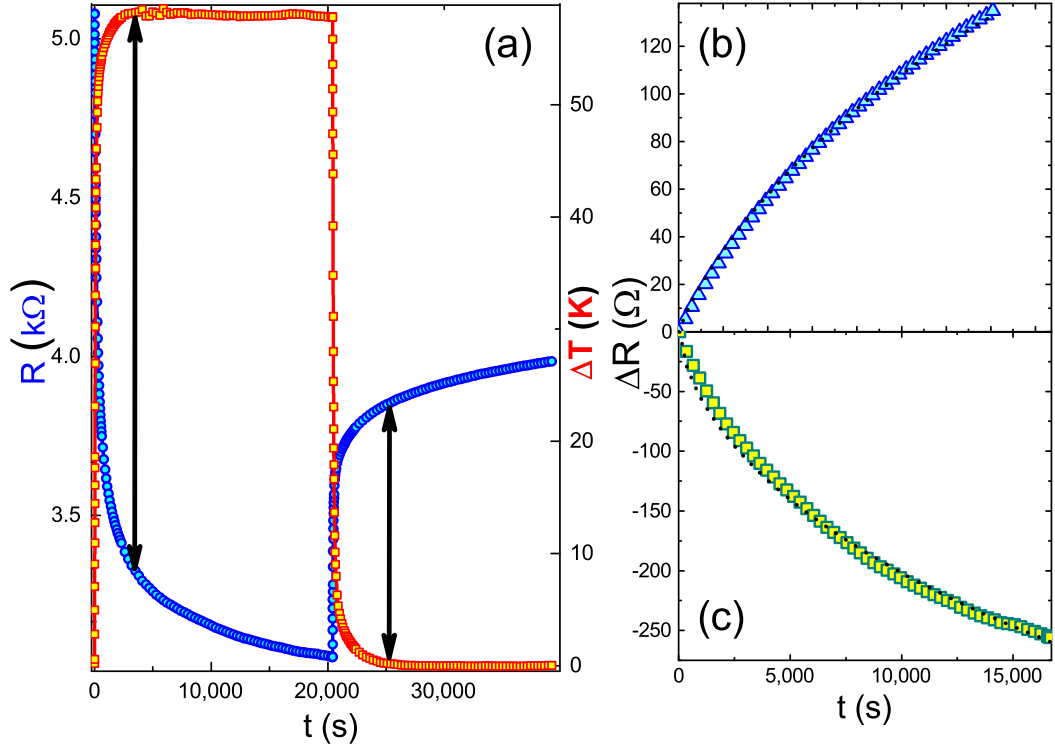}%
\caption{A typical protocol used in thermally-treating the In$_{\text{x}}$O
films. (a) Resistance data R(t) shown in open circles and refer to the left
scale, the sample temperature during the protocol is plotted vs. $\Delta
$T(t)$\equiv$T$_{\text{A}}$(t)-T$_{\text{RT}}$ (T$_{\text{RT}}$=298$\pm$1K)
with open squares and refer to the right scale. The arrows mark the onset of
the constant-temperature time-intervals starting from which fits are shown
(dashed lines) to the $\Delta$R(t) plots in (b) and (c). The data in (c) are
for the annealing-period (circles), and for the data in (b) (triangles) are
for the "swelling" period. The fits (dashed lines) are based on Eq.2 and Eq.3
with the parameters $\beta$=0.84, 0.60 and $\tau$=2x10$^{\text{4}}$s,
6x10$^{\text{4}}$s respectively. }%
\end{figure}

Both $\Delta$R$_{\text{swell}}$(t) and $\Delta$R$_{\text{anneal}}$(t) reflect
changes in the system volume (rarefaction and densification respectively). The
change in volume in the process of thermally-treating In$_{\text{x}}$O films
was demonstrated in an interference experiment using grazing-angle x-ray
technique \cite{15}. The volume change was further correlated with in-situ
resistivity and optical-spectroscopy measurements \cite{15}. It was argued
\cite{14} that the time-dependent processes that occur while the temperature
is constant can be qualitatively accounted for by assuming an effective
two-body potential of the form depicted in Fig.2. Moreover, the similarity of
these effects with those produced by application of pressure could be
explained on the same footing as in thermally-treating the system \cite{14}.
For completeness, we give below a concise summary of these arguments.%
\begin{figure}[ptb]%
\centering
\includegraphics[
height=4.657in,
width=5.5486in
]%
{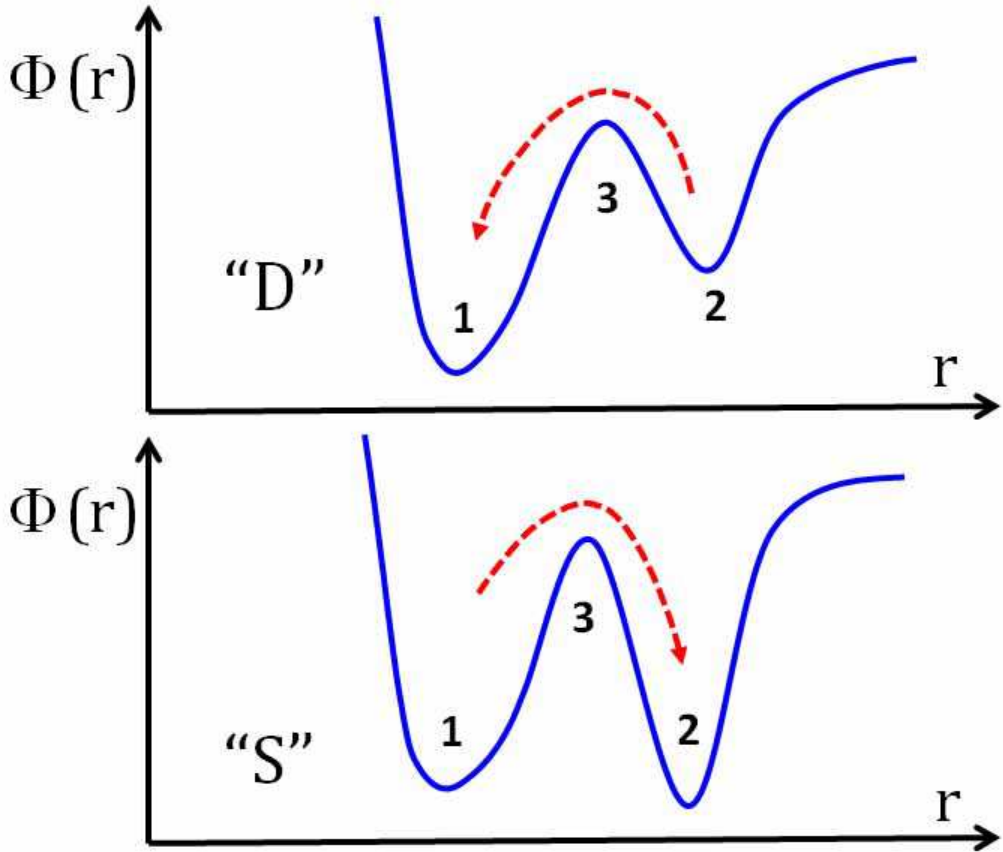}%
\caption{A schematic description of the effective interparticle-potential
$\Phi$ versus their separation r. Two forms of this potential are shown as
representatives of an assumed continuous distribution.}%
\end{figure}

\subsection{A heuristic picture for densification-rarefaction}

The picture is based on the inter-particle potential schematically shown in
Fig.2. The figure depicts two local configurations of the interparticle
potential; "S" and "D" are specific two-state-systems featuring two local
minima. The state labeled S (for `swell') favors a larger interparticle
separation while D favors a denser structure. The system density at a given
temperature and pressure is determined by the values of $\Phi_{i}$'s.
Transitions of the type S,D(1$\rightarrow$2), S,D(2$\rightarrow$1), are
assumed to be controlled by a Boltzmann factor so the transition-rate $\gamma$ is:%

\begin{equation}
\gamma\text{=}\omega\text{\textperiodcentered exp[-}\delta\text{/k}_{\text{B}%
}\text{T]}%
\end{equation}
where $\delta$ is $\delta_{\text{3,1}}$=$\Phi$(3)-$\Phi$(1), $\delta
_{\text{3,2}}$=$\Phi$(3)-$\Phi$(2) respectively, $\omega\approx$%
10$^{\text{12}}$s$^{\text{-1}}$ is the attempt-frequency and T is the temperature.

Many of the local configurations in the as-prepared In$_{\text{x}}$O films,
are probably of the `swell-type' because the samples were quench-condensed
from the vapor phase onto room-temperature substrates and therefore are
similar to rapidly-chilled glasses \cite{1}. Accordingly, S-configurations may
initially be preponderant in the system. When $\Delta$T%
$>$%
0 is applied the balance of occupation in the S(1) and S(2) states changes and
the density will increase towards the level dictated by Boltzmann statistics
and controlled by the distribution of the $\delta_{\text{3,2}}$ barriers. If,
while $\Delta$T is on, {\small t}here are no irreversible structural changes
then the density will eventually saturate at the `equilibrium' value set by
the temperature. In this case the density will acquire its pristine value when
$\Delta$T is reduced to zero. Reversible changes that apparently occur when
the asymptotic value of the resistance is lower than the starting value
signify transformation of S-configurations into D-configurations.

\subsection{Raman spectra}

Amorphous systems are as a rule disordered solids. One of the characteristic
features of these systems is the bosom-peak. This peculiar feature may be
detected in inelastic neutron scattering where it shows up as an excess of
vibration-states over the parabolic Debye spectrum \cite{18,19,20,21,22}. This
boson-peak is routinely observed in Raman scattering experiments where it
exhibits a characteristic asymmetric shape with a peak value around 40-100
cm$^{\text{-1}}$. The effect of pressure on the boson-peak has been studied in
number of disordered systems \cite{22} and showed a systematic reduction of
the peak magnitude with increasing pressure. In this work we show Raman
spectra of before and after thermal-treatment (Fig.3). The figure reveals the
same trend as in other disordered systems; the magnitude of the boson-peak is
significantly smaller following densification. This gives further support to
the observation that thermal-treatment of In$_{\text{x}}$O has qualitatively
the same effect as applying pressure on other structural glasses in agreement
with the heuristic picture \cite{14}. A reduction of the boson-peak magnitude
was also observed after thermally-treating the In$_{\text{x}}$O version used
in \cite{14} but the change relative to the change in the resistance was
smaller. It should also be remarked that with the present set-up, having a
cut-off below 70cm$^{\text{-1}}$, we cannot determine the position where the
boson-peak reaches its maximum value* and therefore we are not able to see if
and by how much it is shifted with annealing.%

\begin{figure}[ptb]%
\centering
\includegraphics[
height=3.9435in,
width=5.5486in
]%
{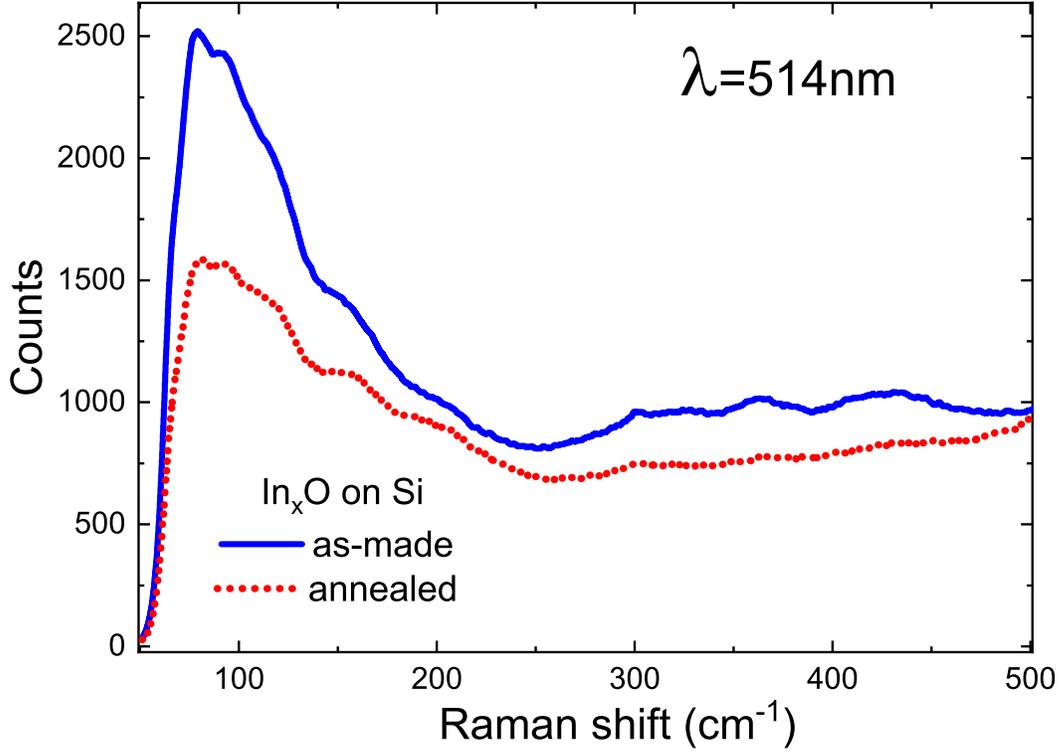}%
\caption{Raman spectra for a 51nm In$_{\text{x}}$O film taken before (full
line) and after (dashed line) thermal-treatment cycle that changed the as-made
sample resistance from 3.1k$\Omega$ to the annealed value of 1.15k$\Omega$.
The two spectra were taken with the same conditions and with a laser power of
7mW.}%
\end{figure}

Further reduction in the boson-peak magnitude is achieved upon crystallizing
the In$_{\text{x}}$O film. Crystallization proceeds rapidly by subjecting the
In$_{\text{x}}$O film to temperatures in excess of $\approx$370K. The Raman
spectrum of the resulting In$_{\text{2}}$O$_{\text{3-x}}$ polycrystalline film
is compared with that of the as-deposited amorphous film and with the spectrum
of the deposition-source material In$_{\text{2}}$O$_{\text{3}}$.%
\begin{figure}[ptb]%
\centering
\includegraphics[
height=3.8588in,
width=5.5486in
]%
{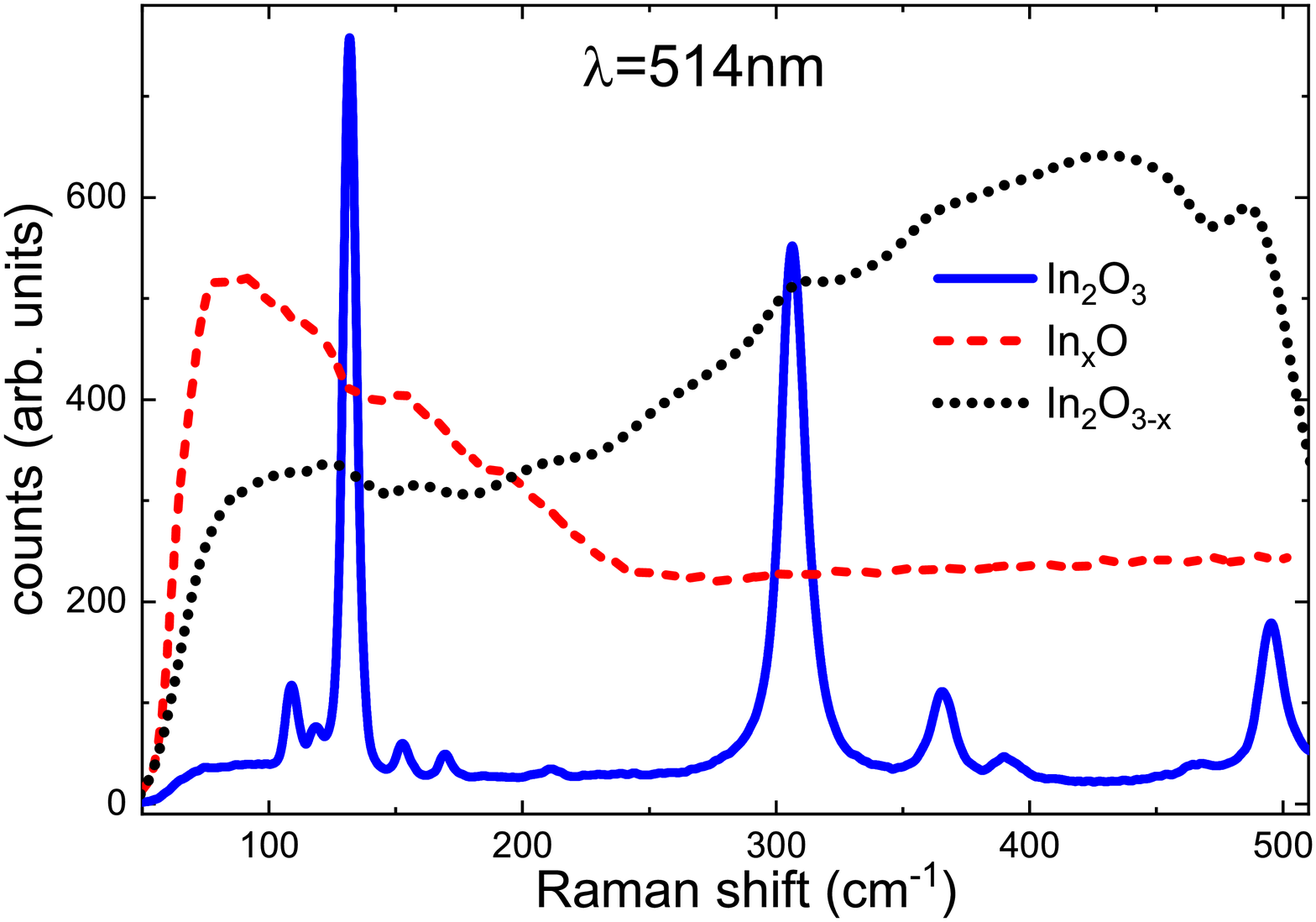}%
\caption{Raman spectra of the three structures of indium-oxide. The plot
labeled as In$_{\text{2}}$O$_{\text{3}}$ (full-line) is based on data taken
from the source material used here for depositing the In$_{\text{x}}$O
(amorphous) 91nm film (dashed line). The latter was later crystallized and
held at 700K for 25 minutes to form In$_{\text{2}}$O$_{\text{3-x}}$ the
polycrystalline version of indium-oxide (dotted-line).}%
\end{figure}

Obviously, Raman spectra for the various structures of indium-oxide samples
are distinctly different. This makes Raman scattering a useful tool to
identify them and, in particular, to distinguish between the oxygen-deficient
(and electron-rich) In$_{\text{2}}$O$_{\text{3-x}}$ and the stoichiometric
compound In$_{\text{2}}$O$_{\text{3}}$. Note that, over the spectral range
studied, some of the expected In$_{\text{2}}$O$_{\text{3}}$ vibration-modes at
109, 135, 307, 366, and 495 cm$^{\text{-1}}$ clearly seen in the
source-material spectrum, are masked or shifted in energy in the
In$_{\text{2}}$O$_{\text{3-x}}$ compound (Fig.4). This presumably is due to
the presence of $\approx$10$^{\text{20}}$cm$^{\text{-3}}$electrons in the
non-stoichiometric system. We found out that Raman spectroscopy is also
sensitive enough to detect a residual amorphous phase in samples that
undergone incomplete crystallization; this shows up as a prominent boson-peak
superimposed on spectrum that looks like that of In$_{\text{2}}$%
O$_{\text{3-x}}$. The presence of a similar amount of amorphous phase on the
same (mostly crystalline) sample was difficult to ascertain by using x-ray or
electron-diffraction techniques as both usually give some signal due to
inelastic scattering that is hard to distinguish from the presence of
amorphous phase.

\subsection{Dynamics of the iso-thermal processes}

We turn now to discuss the slow processes that occur during the iso-thermal
parts of the annealing cycle. The behavior shown in Fig.1 is typical but the
relative magnitude of changes that presumably take place while the system
temperature is constant depend in an intricate way on T$_{\text{A}}$, time of
annealing, and on history. Two examples for protocols that are usually avoided
are shown in Fig.5 and Fig.6.%
\begin{figure}[ptb]%
\centering
\includegraphics[
height=3.7576in,
width=5.5486in
]%
{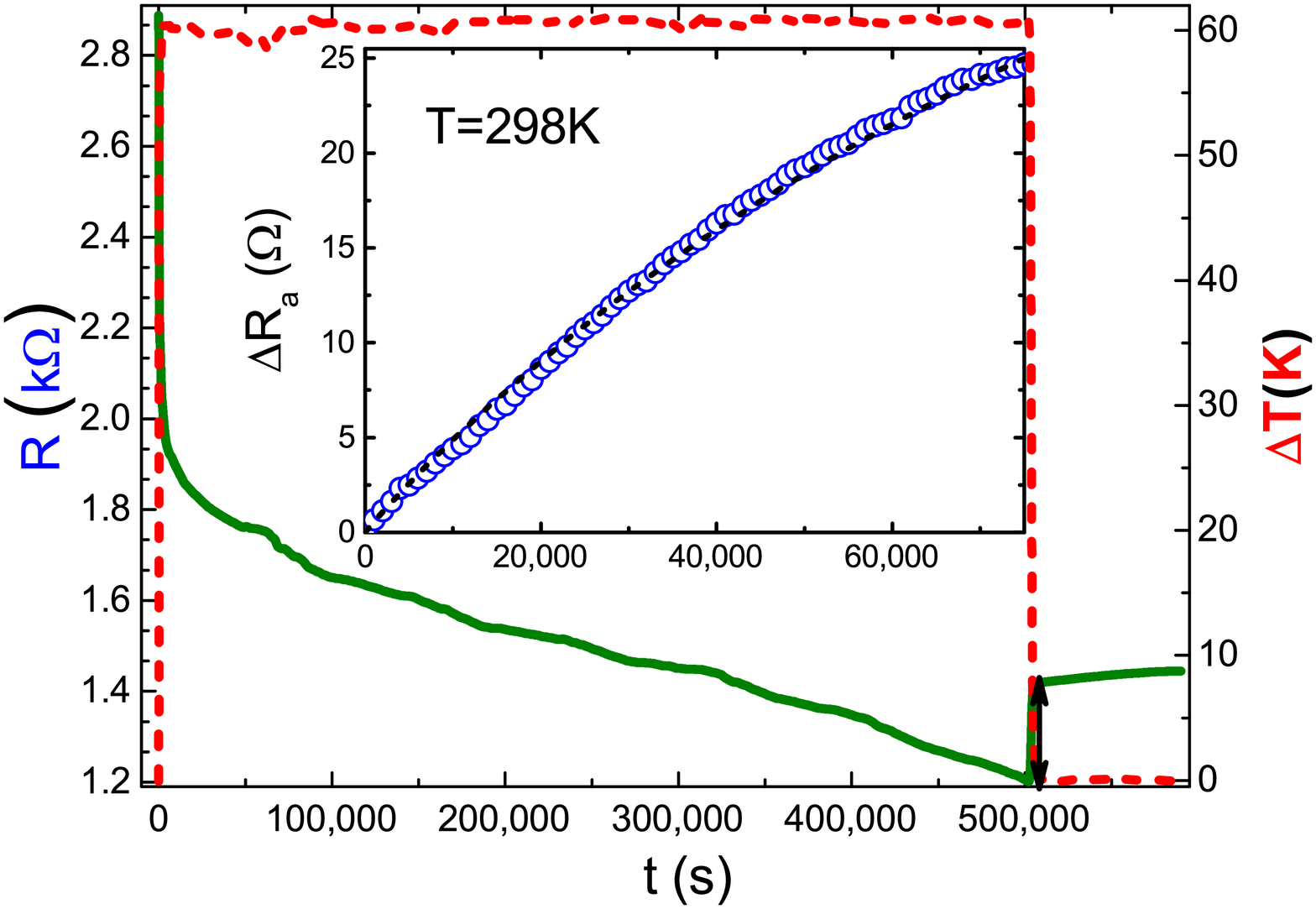}%
\caption{Thermal-annealing cycle carried over an extended period of time. The
sample is a 51nm In$_{\text{x}}$O film (see Fig.1 for details). The inset
shows the swelling-effect reflected in the increase of R(t) of the sample
fitted to Eq.3 with $\beta$=0.96 and $\tau$=10$^{\text{5}}$s (dashed line).
The arrow marks the chosen origin of time for $\Delta$R$_{\text{swell}}$(t).}%
\end{figure}

Figure 5 shows an example for an unusually long heating process which is the
way to get a substantial densification and lowering of the sample resistance
while using moderate annealing temperature. Figure 6, on the other hand, shows
the unwanted result of actually ending up with a \textit{higher} resistance
than the value one started the protocol with. Such an outcome is not rare; it
usually occurs when T$_{\text{A}}$ is close to the temperature the sample was
exposed to previously (for example, during deposition), annealed for short
period of time, or both. In essence, even though the cooling-back to
room-temperatures extends over hundreds of seconds, it is effectively
tantamount to quench-cooling the sample because the cooling-time is still much
shorter than the relaxation-time of the medium. In terms of the heuristic
model, the most relevant configurations (see Fig.2) that are involved in the
swelling effect are those having $\Phi$(1)-$\Phi$(2)$\ $just larger than
k$_{\text{B}}$T$_{\text{A}}$.

An intriguing question to which we have yet no answer is whether it is
possible by careful annealing to completely eliminate `S-type' configurations
from the amorphous system.
\begin{figure}[ptb]%
\centering
\includegraphics[
height=3.6798in,
width=5.5486in
]%
{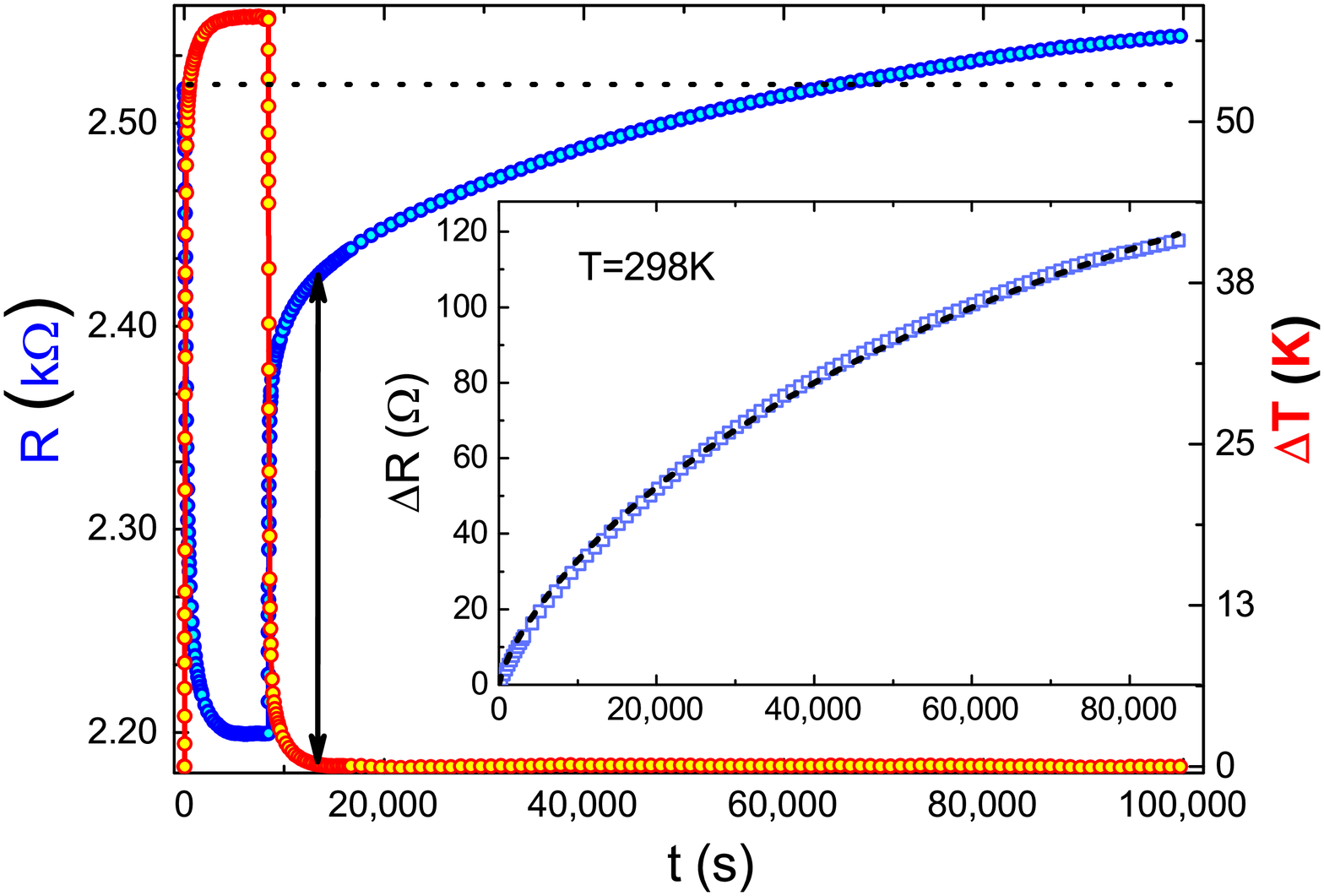}%
\caption{Thermal-treatment cycle for a 51nm In$_{\text{x}}$O film after it was
previously subjected to a thermal cycle at the $\Delta$T=65K for 15,000s. Note
that the resistance at the end of the annealing-cycle flows to a higher value
than the initial value (marked by the dotted line). The inset shows the
swelling effect of the sample fitted (dashed-line) to Eq.3 with $\beta$=0.75
and $\tau$=9.8x10$^{\text{4}}$s (dashed line). The arrow marks the chosen
origin of time for $\Delta$R$_{\text{swell}}$(t).}%
\end{figure}

The temporal dependence of R, while the system is under the
annealing-temperature T$_{\text{A}}$, may be fitted to a stretched-exponential law:%

\begin{equation}
\Delta\text{R}_{\text{anneal}}\text{(t)}=\Delta\text{R}_{\text{a}}%
\exp\text{[-(t/}\tau\text{)}^{\beta}\text{]}%
\end{equation}
with the three parameters $\beta$, $\tau$, and $\Delta$R$_{\text{a}}$.

A complementary behavior takes place during the asymptotic region of the
annealing protocol, where the temperature has settled back at T$_{\text{RT}}$.
In this period R(t) may also be described by a stretched-exponential law (in
general, with different parameters):
\begin{equation}
\Delta\text{R}_{\text{swell}}\text{(t)}=\Delta\text{R}_{\text{s}}%
\text{\{1-}\exp\text{[-(t/}\tau\text{)}^{\beta}\text{]\}}%
\end{equation}

Stretched-exponential time dependence is often observed in the dynamics of
disordered systems \cite{23,24}. There were several attempts to come up with a
microscopic model that accounts for the origin of this so called
Kohlrausch-law \cite{25}. However for discussing the current results a simple
interpretation suggesting that the stretched-exponential is just a weighted
sum of simple exponentials may suffice \cite{26}. In this approach, the
stretched-exponential law is a convoluted effect of parallel events with
relaxation-rates $\gamma$ distributed over a range with probability P($\gamma
$), the parameter $\beta$ is a logarithmic measure of the distribution-width
and $\tau$ is a characteristic relaxation time \cite{26}. We shall argue
however that, unless $\beta\simeq$1, associating $\tau$ with a relaxation time
may convey only part of the characteristics of the dynamics.

Fitting data of a rather plain form such as our $\Delta$R$_{\text{anneal}}$(t)
or $\Delta$R$_{\text{swell}}$(t) plots to Eq.2 or Eq.3, involving 3
parameters, should be taken with some reservations. The main problem in trying
to fit such data to Eq.2 or Eq.3 is the time it takes the sample temperature
to stabilize upon heating-up to T$_{\text{A}}$ and after cooling-back to
room-temperature. This introduces uncertainty in the assignment of the origin
of time. Nevertheless, fitting data to these stretched-exponential expressions
is still a reasonable tool to estimate (by extrapolation) the asymptotic value
of the resistance, and to compare changes in dynamics on a relative basis. Our
choice of the "starting-time" for the iso-thermal process is marked by the
arrows in figures 1, 5, and 6.

\subsection{How disorder is reflected in the swelling dynamics}

Next we wish to focus on how the dynamics observable through analyzing
$\Delta$R$_{\text{swell}}$(t) data reflect on the degree of the disorder after
a given thermal-treatment cycle. In transport measurements, it is the sample
resistivity $\rho_{\text{RT}}$ that is often taken as the criterion for
disorder. Thermal-annealing modifies the sample resistance at the end of the
heat-cycle and therefore it changes the disorder. However, in addition to the
value of the average resistivity, there are changes observable in the dynamics
of the `after-effect' that may give further information on the modified system disorder.

We have studied three series of samples that were progressively annealed by
subjecting them to thermal-treatment for successively longer times, higher
T$_{\text{A}}$ (or both). This resulted in a lower resistance at the end of
each annealing cycle. The dynamics of the system was then tested at each
cycle-end by the very same protocol: The sample was exposed to T$_{\text{A}}%
$=325$\pm$1K for $\approx$60 minutes, then allowed to cool back to
room-temperature and the ensuing $\Delta$R$_{\text{swell}}$(t) was monitored.
A fit was then made for these data to Eq.3 using the same criteria for the
choice of the origin of time for each heat-treatment cycle. It turns out that
the best-fits parameters $\beta$ and $\tau$ consistently increased as the
annealing stages progressed and the sample resistance decreased. Three plots
that illustrate this trend are shown in fig.7. Note that the asymptotic value
of $\Delta$R$_{\text{swell}}$ gets progressively smaller with further
thermal-cycles but it seems to reach a limit as further annealing gives little
or no improvement of the conductance. In addition, the time dependence of
$\Delta$R$_{\text{swell}}$(t) also changes: More annealing causes $\Delta
$R$_{\text{swell}}$(t) to exhibit a smaller curvature. This is illustrated in
Fig.7 comparing the plots for the 1st and 3rd thermal-treatment cycles (the
latter is also shown blown-up to facilitate comparison of curvature).%

\begin{figure}[ptb]%
\centering
\includegraphics[
height=3.928in,
width=5.5486in
]%
{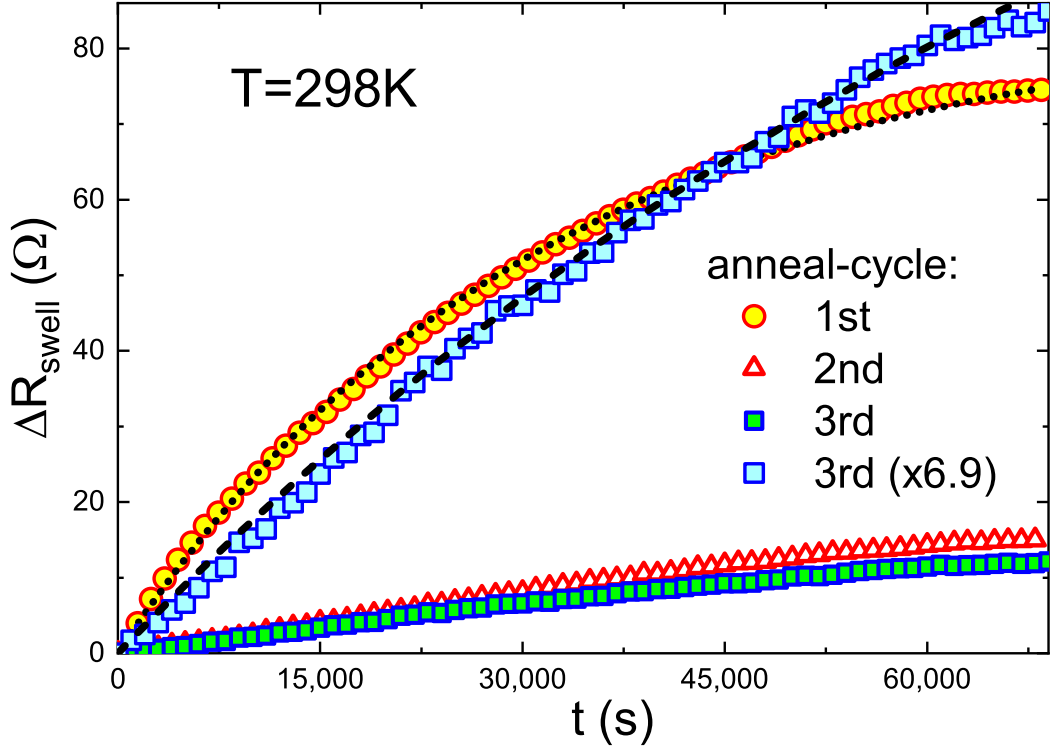}%
\caption{The time dependent $\Delta$R$_{\text{swell}}$(t) for three
consecutive thermal-annealing cycles on the same 51nm In$_{\text{x}}$O film
where the sample was held at the same $\Delta$T and for the same time then
cooled back to room-temperature and its R(t) was monitored. These $\Delta
$R$_{\text{swell}}$(t) data are fitted (dashed line) to Eq.3 with $\beta
$=0.51, 0.88, 0.96 and $\tau$=1.6x10$^{\text{4}}$s, 6x10$^{\text{4}}$s,
8.4x10$^{\text{4}}$s for thermal-cycle \#1, 2, and 3 respectively.}%
\end{figure}

The $\Delta$R$_{\text{swell}}$(t) plots in Fig.7 are results of testing three
consecutive annealing-stages yielding best-fit values for $\beta$ and for the
relaxation-times $\tau$ that are listed in the table below. These were
purposefully chosen for display because their $\beta^{\prime}$s are close to
1/2, 7/8 and 24/25 for thermal-cycle 1, 2, and 3 respectively. Note that
P($\gamma$,$\beta$), where $\beta$ equals ratio of integers, may be readily
calculated by the prescription suggested by Johnston \cite{26}.

\begin{center}%
\begin{tabular}
[c]{|c|c|c|}\hline
annealing cycle \# & $\beta$ & $\tau$ (x10$^{\text{4}}$s)\\\hline
1 & 0.51$\pm$0.05 & 1.6$\pm$0.08\\\hline
2 & 0.87$\pm$0.02 & 6.4$\pm$0.1\\\hline
3 & 0.96$\pm$0.015 & 8.1$\pm$0.2\\\hline
\end{tabular}

\end{center}

These specific rate-distributions are plotted in Fig.8a as function of the
dimensionless variable $\gamma\tau$ \cite{26}.
\begin{figure}[ptb]%
\centering
\includegraphics[
height=3.6495in,
width=5.5486in
]%
{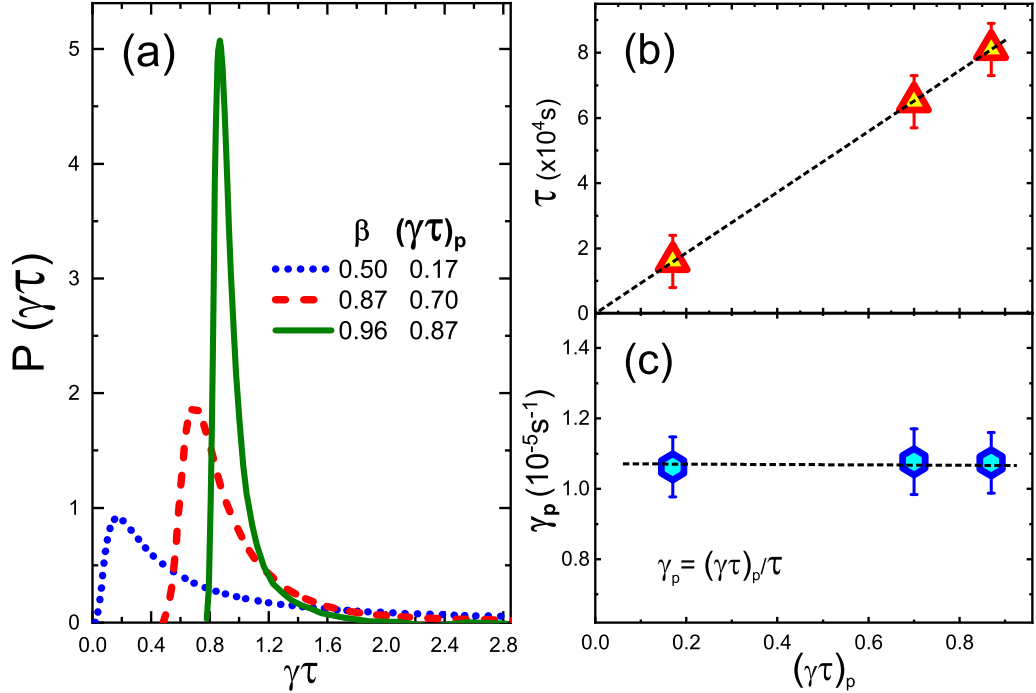}%
\caption{(a) The normalized probability-distribution of the dimensionless
parameter $\gamma t$ that gives the stretched exponential relaxation law
\cite{26} for three values of $\beta$. ($\gamma t$)$_{\text{p}}$ is the value
at the respective point where $\partial$P($\gamma t$)/$\partial$($\gamma
t$)=0. (b) The best fitted value of the parameter $\tau$ for the three
annealing cycles shown in Fig.7. (c) The transition rates of these samples at
the peak of the respective P($\gamma t$) distribution.}%
\end{figure}

It would be natural to take the value of the fitted $\tau$ as a characteristic
relaxation-time for a given sample. In this case one may conclude from Fig.8b
that the dynamics of the swelling effect becomes slower with progressive
annealing. However, plotted against the peak-value ($\gamma\tau$)$_{\text{p}}$
of the P($\gamma$) associated with the best-fitted $\beta$, one observes an
intriguing correlation in our data; $\tau$ is linear with ($\gamma\tau
$)$_{\text{p}}~$(Fig.8b)$.$This, in turn, means that $\gamma_{\text{p}}$, the
most probable rate involved in the dynamics of $\Delta$R$_{\text{swell}}$(t)
appears to remain constant\textit{ }as the annealing proceeds to further
reduce the sample resistance (Fig.8c). It should be emphasized that the trend
of $\beta$ getting larger with further annealing has been observed in
\textit{all} our thermal-treatment experiments, but not knowing the associated
P($\gamma t$) we cannot ascertain that the $\tau\propto$($\gamma\tau
$)$_{\text{p}}$ relation holds in each interim annealing stage. What is clear
is that larger $\beta$ means narrower, more symmetric rate-distribution.
Therefore it is fair to conclude that, in addition to increasing the material
density, thermally-treating In$_{\text{x}}$O films improves their homogeneity.
This agrees with independent data obtained by using x-ray interferometry
\cite{15}. The lesson here is that the value of the best-fitted $\tau$ by
itself conveys only part of the information relevant for the system dynamics
(unless $\beta$ is close to unity). When $\beta$ is small the relaxation-time
is an ill-defined quantity and its value may depend on the specific protocol
one uses to estimate it.

Using Eq.1 the effective barrier that would give the most probable
transition-rate found above for the In$_{\text{x}}$O films may be estimated
by:%
\begin{equation}
\delta_{\text{p}}=\text{k}_{\text{B}}\text{T}\ln\left(  \frac{\omega}%
{\gamma_{\text{p}}}\right)
\end{equation}

With k$_{\text{B}}$T$\simeq$25meV, $\omega\simeq$10$^{\text{12}}$%
s$^{\text{-1}}$, and $\gamma_{\text{p}}\simeq$10$^{\text{-5}}$s$^{\text{-1}}$
(from Fig.8c) one gets $\delta_{\text{p}}\simeq$1.0$\pm$0.05~eV. It should be
noted that, within the experimental error, this $\delta_{\text{p}}$ is
identical with the value estimated for the barrier of another version of this
material with a lower carrier-concentration \cite{14}. The order of magnitude
difference (in terms of carrier-concentration) between these versions of
In$_{\text{x}}$O has apparently little effect on the value of the barrier.
This suggests that its origin has to do with local chemistry rather than the
specific stoichiometry of the compound.

With a barrier of the order of 1eV the only dynamics expected below $\approx
$100K on a time scale relevant for most experiments would be of quantum
nature; at liquid nitrogen temperatures the over-the-barrier transition-rate
would already be smaller than 10$^{\text{-54}}$s$^{\text{-1}}$.

The lack of observable increase with time of the average resistance for a
sample immersed in liquid helium suggests that tunneling-transition is also
practically suppressed. The effective distance for tunneling may be of atomic
dimension but it probably involves many-body effects making the mass of the
tunneling object too large. This effective freezing-out of the system disorder
makes thermal-treatment a useful tool for a variety of low-temperature studies
where one seeks a way to fine-tune the disorder. Not only does it allow to
change the average disorder over a range large enough to affect the
metal-insulator or superconductor insulator transition, but by letting the
system retrieve some disorder through the swelling-effect, it also allows a
certain degree of reversibility. This is especially useful for low temperature
study of the insulating phase where a tiny change in room-temperature
resistance translates into an exponentially large change at the measuring
temperatures \cite{15}.

\section{Summary}

We have studied in this work several effects associated with
thermally-treating \-In$_{\text{x}}$O films. Two effects that were studied
previously are a reduction in the system volume and optical-gap during
heat-treatment as well as a partial recovery of these quantities after the
sample is cooled back to room-temperature \cite{14}. Here it is shown that
thermal-treatment also leads to a reduction of the boson-peak magnitude. These
three effects were observed in other disordered systems by applying pressure.
The response of a system volume to application of pressure and the rarefaction
that takes place once pressure is relieved appeals to one's intuition based on
everyday experience. That thermal-treatment has similar effects may seem
counter-intuitive, again, due to common experience; raising temperature
usually (but not always \cite{27}) causes system to expand. It is in the
nonequilibrium solids like amorphous systems that this peculiar behavior may
turn out to be prevalent. There is no reason to believe that In$_{\text{x}}$O
films are unique in this regard.

Monitoring the system resistance allow convenient and sensitive tracking of
the dynamics and energetics associated with these phenomena during both
densification and after the heat-treatment is terminated and the system
relaxes towards a new (metastable) state. These processes are\ presumably
activated and exhibit temporal dependencies that fit the Kohlrausch-law often
encountered in experiments involving slow-dynamics of structural glasses. The
parameters that fit the relaxation-law $\beta$ and $\tau$ change
systematically with the degree of annealing. An analysis of these changes
suggest that, in addition to densification, there is a narrowing of the
rate-distribution. It would be interesting to see how general are these
findings by studying the effects of thermal-treatment on other metallic glasses.

The assistance by Dr. Anna Radko with the Raman spectra work is gratefully
acknowledged. This research has been supported by the 1030/16 grant
administered by the Israel Academy for Sciences and Humanities.

*After this work was published, we have found that the maximum magnitude of
the boson-peak in this version of In$_{\text{x}}$O occurs at 38$\pm$2
cm$^{\text{-1}}$ and does not shift with thermal-treatment (work performed in
collaboration with Itai Bet-hazavdi and Ilana Bar from Ben-Gurion University
of the Negev).


\begin{thebibliography}{99}                                                                                               %


\bibitem {1}P. W. Bridgman and I. \v{S}imon, J. of Appl. Phys., \textbf{24},
405 (1953).

\bibitem {2}S. Sakka and J. D. Mackenzie, J. of Non-Crys Sol., \textbf{1}, 107 (1969).

\bibitem {3}J. D. Mackenzie, J. of Am. Ceramic Soc., \textbf{46}. 461 (1963).

\bibitem {4}J. D. Mackenzie, J. of Am. Ceramic Soc., \textbf{46}, 470 (1963).

\bibitem {5}Wu and H.L. Luo, Journal of Non-Crystalline Solids \textbf{18,} 21 (1975).

\bibitem {6}N. Sakai and H. Fritzsche, Phys. Rev. B \textbf{15}, 973 (1977).

\bibitem {7}Seinosuke Onari, Takao Inokuma, Hiromichi Kataura, and Toshihiro
Arai, Phys. Rev. B \textbf{35,} 4373 (1987).

\bibitem {8}A. Polian and M. Grimsditch, Phys. Rev. B \textbf{41}, 6086 (1990).

\bibitem {9}S. Susman, K. J. Volin, D. L. Price, M. Grimsditch, J. P. Rino, R.
K. Kalia, and P. Vashishta, G. Gwanmesia, Y. Wang, and R. C.
Liebermann\textbf{,} Phys. Rev. B \textbf{43, }1194 (1991).

\bibitem {10}Norio Ookubo, Yasuhiro Matsuda, and Noritaka Kuroda, Applied
Physics Lett., \textbf{63}, 346 (1993).

\bibitem {11}Daniel J. Lacks, Phys. Rev. Lett., \textbf{30}, 5385 (1998).

\bibitem {12}K. Miyauchi, J. Qiu, M. Shojiya, Y. Kawamoto, N.
Kitamura\textbf{, }J. of Non-Crys Sol., \textbf{279, }186 (2001).

\bibitem {13}V. V. Brazhkin, E. Bychkov, and O. B. Tsiok, Phys. Rev. B
\textbf{95}, 054205 (2017).

\bibitem {14}Z. Ovadyahu, Phys. Rev. B \textbf{95}, 214207 (2017).

\bibitem {15}Z. Ovadyahu , Phys. Rev. B. \textbf{95}, 134203 (2017).

\bibitem {16}D. Shahar and Z. Ovadyahu, Phys. Rev. B \textbf{46}, 10917
(1992); D. Kowal and Z. Ovadyahu, Solid State Comm., \textbf{90}, 783 (1994);
V. Gantmakher, International Journal of Modern Physics B, \textbf{12}, Nos.
29, 30 \& 31 (1998); G. Sambandamurthy, L. W. Engel, A. Johansson, and D.
Shahar, Phys. Rev. Lett. \textbf{92}, 107005 (2004); M. A. Steiner, G.
Boebinger, and A. Kapitulnik, Phys. Rev. Lett. \textbf{94}, 107008 (2005); M
Steiner, A Kapitulnik, Physica C, \textbf{422}, 16 (2005); Myles A. Steiner,
Nicholas P. Breznay, and Aharon Kapitulnik, Phys. Rev. B \textbf{77}, 212501
(2008); Nicholas P. Breznay, Myles A. Steiner, Steven Allan Kivelson, and
Aharon Kapitulnik, PNAS, \textbf{113}, 215 (2015); P. Breznay and Aharon
Kapitulnik, Science advances, \textbf{3}, e1700612 (2017); T. I. Baturina, A.
Bilu\v{s}i\'{c}, A.Yu. Mironov, V. M. Vinokur, M. R. Baklanov, and C. Strunk,
Physica C \textbf{468,} 316 (2008); S. Poran, E. Shimshoni, and A. Frydman,
Phys. Rev. B \textbf{84}, 014529 (2011); D. Sherman, G. Kopnov, D. Shahar, and
A. Frydman, Phys. Rev. Lett. \textbf{108}, 177006 (2012); Yeonbae Lee, Aviad
Frydman, Tianran Chen, Brian Skinner, and A. M. Goldman, Phys. Rev. B
\textbf{88}, 024509 (2013); Benjamin Sac\'{e}p\'{e}, Thomas Dubouchet, Claude
Chapelier, Marc Sanquer, Maoz Ovadia, Dan Shahar, Mikhail Feigel'man and Lev
Ioffe, Nature Physics, \textbf{7}, 239 (2011); Daniel Sherman, Uwe S. Pracht,
Boris Gorshunov, Shachaf Poran, John Jesudasan, Madhavi Chand, Pratap
Raychaudhuri, Mason Swanson, Nandini Trivedi, Assa Auerbach, Marc Scheffler,
Aviad Frydman \& Martin Dressel, Nature Physics \textbf{11}, 1882 (2015);
Ilana M. Percher, Irina Volotsenko, Aviad Frydman, Boris I. Shklovskii, and
Allen M. Goldman, Phys. Rev. B \textbf{96}, 224511 (2017).

\bibitem {17}U. Givan and Z. Ovadyahu, Phys. Rev. B \textbf{86}, 165101 (2012).

\bibitem {18}U. Strom, and P. C. Taylor, Phys. Rev. B, \textbf{16} 5512 (1977).

\bibitem {19}U. Buchenau, M. Prager, N. N\"{u}cker et al., Phys. Rev. B,
\textbf{34} 5665 (1986).

\bibitem {20}V. K. Malnikovsky, V. N. Novikov, P. P. Parshin, A. P. Sokolov
and M. G. Zemlyanov, Europhys. Lett., \textbf{11} 43 (1990).

\bibitem {21}Walter Schirmacher, Gregor Diezemann, and Carl Ganter, Phys. Rev.
Lett. \textbf{81}, 136 (1998).

\bibitem {22}K. Niss, B. Begen, B. Frick, J. Ollivier, A. Beraud, A. Sokolov,
V. N. Novikov, and C. Alba-Simionesco, Phys. Rev. Lett. \textbf{99}, 055502
(2007); L. Hong, B. Begen, A. Kisliuk, C. Alba-Simionesco, V. N. Novikov, and
A. P. Sokolov, Phys. Rev. B \textbf{78}, 134201 (2008); V. L. Gurevich, D. A.
Parshin, and H. R. Schober, Phys. Rev. B \textbf{71}, 014209 (2005); S. Sugai
and A. Onodera, Phys. Rev. Lett. \textbf{77}, 4210 (1996); A. Monaco, A. I.
Chumakov, G. Monaco, W. A. Crichton, A. Meyer, L. Comez, D. Fioretto, J.
Korecki, and R. R\"{u}ffer, Phys. Rev. Lett. \textbf{97}, 135501 (2006); L.
Hong, B. Begen, A. Kisliuk, S. Pawlus, M. Paluch, and A. P. Sokolov, Phys.
Rev. Lett. \textbf{102}, 145502 (2009); H. R. Schober, U. Buchenau, and V. L.
Gurevich, Phys. Rev. B \textbf{89}, 014204 (2014); Hiroshi Shintani and Hajime
Tanaka, nature materials \textbf{7}, 870 (2008).

\bibitem {23}C. A. Angell, K. L. Ngai, G. B. McKenna, P. F. McMillan, and S.
W. Martin, J. of Appl. Phys., \textbf{88}, 3113 (2000).

\bibitem {24}Bingyu Cui, Rico Milkus, and Alessio Zaccone, Phys. Rev. E
\textbf{95}, 022603, (2017).

\bibitem {25}E. W. Montroll, J. T. Bendler, J. Stat. Phys. \textbf{34},129
(1984); R. G. Palmer, D. L. Stein, E. Abrahams, and P. W. Anderson, Phys. Rev.
Lett., \textbf{53}, 958 (1984); J. S. Langer, S. Mukhopadhyay, Phys. Rev. E
\textbf{77}, 061505 (2008); J. C. Phillips, Rep. Prog. Phys. \textbf{59}, 1133
(1996); I. M. Lifshitz, Usp. Fiz. Nauk \textbf{83,} 617 (1964) [Engl. trans.
Sov. Phys. Usp. \textbf{7}, 549 (1965)]; R. Friedberg, J. M. Luttinger, Phys.
Rev. B \textbf{12}, 4460 (1975); P. Grassberger, I. Procaccia, J. Chem. Phys.
\textbf{77}, 6281 (1982).

\bibitem {26}D. C. Johnston, Phys. Rev. B \textbf{74}, 184430 (2006).

\bibitem {27}T. A. Mary, J. S. O. Evans, T. Vogt, and A. W. Sleight, Science,
\textbf{272}, 90 (1996); Alexandra K A Prydedag, Kenton D Hammondsdag, Martin
T Dove, Volker Heine, Julian D Gale and Michele C Warren, Journal of Physics:
Condensed Matter, 8, 10973 (1996); John S. O. Evans, J. Chem. Soc., Dalton
Trans., \textbf{19}, 3317 (1999); Andrew L. Goodwin, Mark Calleja, Michael J.
Conterio, Martin T. Dove, John S. O. Evans, David A. Keen, Lars Peters,
Matthew G. Tucker, Science, \textbf{319}, 794 (2008).
\end{thebibliography}
\end{document}